\documentclass[%
 reprint,
 amsmath,amssymb,
 aps,
]{revtex4-1}

\usepackage{graphicx}
\usepackage{dcolumn}
\usepackage{bm}
\usepackage{color}

\begin{document}


\title{Full Coupled-Cluster Reduction \\ for Accurate Description of Strong Electron Correlation}

\author{Enhua Xu}
\author{Motoyuki Uejima}
\author{Seiichiro Lenka Ten-no}
\email{tenno@garnet.kobe-u.ac.jp}
\altaffiliation[Also at ]{Graduate School of System Informatics, Kobe University, Nada-ku, Kobe 657-8501, Japan}
\affiliation{Graduate School of Science, Technology, and Innovation, Kobe University, Nada-ku, Kobe 657-8501, Japan}

\date{\today}

\begin{abstract}
A full coupled-cluster expansion suitable for sparse algebraic operations is developed by expanding the commutators of the Baker-Campbell-Hausdorff series explicitly for cluster operators in binary representations.
A full coupled-cluster reduction that is capable of providing very accurate solutions of the many-body Schr\"odinger equation is then initiated employing screenings to the projection manifold and commutator operations.
The projection manifold is iteratively updated through the single commutators $\left\langle \kappa  \right| [\hat H,\hat T]\left| 0 \right\rangle$ comprised of the primary clusters $\hat T_{\lambda}$ with substantial contribution to the connectivity.
The operation of the commutators is further reduced by introducing a correction, taking into account the so-called exclusion principle violating terms, that provides fast and near-variational convergence in many cases.
\end{abstract}

\maketitle
Approximating the full configuration interaction (FCI) solution for the many-electron electronic Schr\"odinger equation accurately, based on a basis set expansion, is still a challenging task in {\it ab initio} quantum theory in chemistry and physics especially for strongly correlated electronic systems.
One of the most significant advancements in this context is the density matrix renormalization group (DMRG) \cite{white,dmrg}, which has increased the applicability of the FCI wave function approach.
The limitation of the basis set convergence has been transcended by the F12 theory for complex systems \cite{f12}.
Stochastic approaches in configuration space \cite{fciqmc09,i-fciqmc,s-fciqmc,msqmc1} have also been increasing the efficacy as a new means to approximate the FCI solution with reduced memory requirements, which has stimulated the investigation of adaptive CI methods as their deterministic alternatives in recent years \cite{aci,asci,hbci}.
Nevertheless, such approaches based on a linear expansion presented long ago, {\it e.g.} Ref. \onlinecite{cipsi}, necessitate a truncation in the configuration space accompanying a size-inconsistency error, which is difficult to prevent completely once initiated from a truncated CI expansion.

Coupled-cluster (CC) theory \cite{cizek,rod_rev} features the size-extensivity {\it a priori} owing to the exponential wave function ansatz, and has been the most successful framework in {\it ab initio} quantum chemistry for single-reference molecules.
Unfortunately, CC treatments of strongly correlated systems require the inclusion of higher-rank cluster operators within a single-reference framework, and such an implementation permitting very high excitations has been realized so far only with the help of an FCI code \cite{hirata} or automated code synthesis \cite{kallay}.
Therefore, the development of a CC alternative to the adaptive CI has been quite limited to date.
An adaptive coupled-cluster (@CC) approach \cite{acc}, which utilizes an importance selection function has been proposed and tested with the assistance of code synthesis \footnote{D. I. Lyakh, private communication (2018).} for systems where FCI calculations are feasible.
More recently, cluster decomposition of FCI wave functions has been introduced to investigate cluster operators needed for describing strongly correlated systems \cite{cdfci}.
Stochastic adaptations of CC \cite{ccqmc}, coupled-electron pair approximation (CEPA) \cite{msqmc2}, and selecting higher-order clusters \cite{mcme} have also been developed.

In this  Letter, a novel approach for the computation of many-fermionic systems is introduced based on the full coupled-cluster (FCC) expansion.
Systematic reductions are developed in the necessary projection manifold and commutator operations to exploit the sparsity of the FCC wave function.
In this FCC reduction (FCCR), a proper treatment of the so-called exclusion-principle violating (EPV) terms plays an important role to accelerates the convergence towards the exact solution.

What follows is the setup of the FCC expansion: For a given basis set, the exact solution of the $N$-electronic Schr\"odinger equation is expressed either by the linear (FCI) or, equivalently, by the exponential (FCC) ansatz,
\begin{align}
| \Psi \rangle &= (1+\hat C_1+\hat C_2+\hat C_3+ \dots +\hat C_N) |0\rangle \label{eq:fci} \\
&= \exp(\hat T_1+\hat T_2+\hat T_3+ \dots +\hat T_N)|0\rangle, \label{eq:fcc}
\end{align}
where, $\hat C_k$ and $\hat T_k$ denote $k$-fold excitation operators with respect to a suitable Fermi vacuum $|0\rangle$ of a single Slater-determinant.
The dimension of the FCI expansion (\ref{eq:fci}) increases combinatorially with respect to the numbers of electrons and orbitals, while the exponential ansatz (\ref{eq:fcc}) is expected to be a much more compact representation with increasing the system-size, as indicated by the formal relation,
\begin{align}
\hat T &= \ln (1+ \hat C) \nonumber \\
&=\hat C - \hat C^2/2 + \hat C^3/3 - \hat C^4/4 + \dots, \label{eq:fci_fcc}
\end{align}
in which $\hat T$ is exempted from disconnected products for separable correlation events in the FCI expansion.
This compactification was investigated numerically in the cluster decomposition of FCI wave functions \cite{cdfci}.

The standard CC formulae for FCC are used, obtained by the projection of the similarity-transformation of the Schr\"odinger equation onto the projection manifold of the FCI space $\{\langle \kappa | \}$,
\begin{align}
&\langle \kappa | \exp(-\hat T) \hat H \exp(\hat T) | 0 \rangle = \langle \kappa | \hat H | 0 \rangle + {\sum\limits_{\lambda} {\langle \kappa | [\hat H, \hat T_{\lambda}] | 0 \rangle}} \nonumber \\
&+ \frac{1}{2}{\sum\limits_{\lambda\mu} {\langle \kappa | [[\hat H, \hat T_{\lambda}],\hat T_{\mu}] | 0 \rangle}}
+ \frac{1}{6}{\sum\limits_{\lambda\mu\nu} {\langle \kappa | [[[\hat H, \hat T_{\lambda}],\hat T_{\mu}],\hat T_{\nu}] | 0 \rangle}} \nonumber \\
&+ \frac{1}{24}{\sum\limits_{\lambda\mu\nu o} {\langle \kappa | [[[[\hat H, \hat T_{\lambda}],\hat T_{\mu}],\hat T_{\nu}],\hat T_{o}] | 0 \rangle}}=0, \label{eq:fcc_w}
\end{align}
and the state energy,
\begin{align}
E=\langle 0 | \hat H \exp(\hat T) | 0 \rangle,\label{eq:fcc_e}
\end{align}
where the cluster operator is
\begin{align}
\hat T = {\sum\limits_{\lambda} \hat T_{\lambda}} = {\sum\limits_{\lambda} t_{\lambda} \hat a_{\lambda}},
\end{align}
$\kappa,\lambda,\dots$ stand for sets of particle-hole excitation indices, and $t_{\lambda}$ and  $\hat a_{\lambda}$ are cluster amplitudes and the corresponding excitation operators, $\left| \lambda \right\rangle = \hat a_\lambda \left| 0 \right\rangle$, respectively.

The present implementation of FCC employs a direct binary representation of $\hat a_{\lambda}$ for the occupations of the $\alpha$ and $\beta$ strings of the resulting Slater-determinants $| \lambda \rangle$ subject to the nomalization $\langle \kappa | \lambda \rangle = \delta_{\kappa\lambda}$.
The commutators through the quartic order in (\ref{eq:fcc_w}) are computed explicitly for a given set of arguments $\{\kappa, \lambda, \mu, \dots\}$ corresponding to the projection $\langle \kappa |$ and excitations $\hat a_\lambda, \hat a_\mu, \dots$ by expanding the commutators into a sum of Hamiltonian matrix elements over Slater-determinants.
For instance, the expansion
\begin{align}
\langle \kappa | [[[\hat H, \hat a_{\lambda}],\hat a_{\mu}],\hat a_{\nu}] | 0 \rangle =& \langle \kappa | \hat H \hat a_{\lambda} \hat a_{\mu} | \nu \rangle - \langle \kappa | \hat a_{\lambda} \hat H a_{\mu} | \nu \rangle \nonumber \\
- \langle \kappa | \hat a_{\mu} & \hat H \hat a_{\lambda} | \nu \rangle + \langle \kappa | \hat a_{\mu} \hat a_{\lambda} \hat H | \nu \rangle \nonumber \\
- \langle \kappa | \hat a_{\nu} & \hat H \hat a_{\lambda} | \mu \rangle + \langle \kappa | \hat a_{\nu} \hat a_{\lambda} \hat H | \mu \rangle \nonumber \\
+ \langle \kappa | \hat a_{\nu} & \hat a_{\mu} \hat H | \lambda \rangle - \langle \kappa | \hat a_{\nu} \hat a_{\mu} \hat a_{\lambda} \hat H | 0 \rangle, \label{eq:com3}
\end{align}
is performed inside a naive 4-fold loop over $\lambda$, $\mu$, $\nu$, and $\kappa$, and the actions of the excitation and de-excitation operators, $\hat a_{\mu} | \nu \rangle$, $\hat a_{\lambda} | \nu \rangle$, $\hat a_{\lambda} | \mu \rangle$, $\hat a_{\lambda} \hat a_{\mu} | \nu \rangle$, $\hat a_{\lambda}^{\dagger} | \kappa \rangle$, $\hat a_{\mu}^{\dagger} | \kappa \rangle$, $\hat a_{\nu}^{\dagger} | \kappa \rangle$, $ \hat a_{\mu}^{\dagger} \hat a_{\lambda}^{\dagger} | \kappa \rangle$, $ \hat a_{\nu}^{\dagger} \hat a_{\lambda}^{\dagger} | \kappa \rangle$, $ \hat a_{\nu}^{\dagger} \hat a_{\mu}^{\dagger} | \kappa \rangle$, and $ \hat a_{\nu}^{\dagger} \hat a_{\mu}^{\dagger} \hat a_{\lambda}^{\dagger} | \kappa \rangle$, are converted to signed Slater-determinants.
Then, the Hamiltonian matrix elements over the determinants are assembled for (\ref{eq:com3}).
The loop over the projection $\langle \kappa|$ is not a mandatory setup within FCC since only a limited number of $\langle \kappa|$ interact with the contractions of $ [[[\hat H, \hat a_{\lambda}],\hat a_{\mu}],\hat a_{\nu}] | 0 \rangle$.
Nevertheless, this structure is retained keeping the capability of cluster operator selections in mind.
Alternatively, several criteria to rapidly discriminate between a combination of the arguments $\{\kappa, \lambda, \mu, \dots\}$ giving a null result are introduced in a highly parallelizable manner.
The above process is carried out using a numerical library for quantum Monte Carlo calculations in configuration space \cite{msqmc2} containing bitwise operations over Slater-determinants in the binary representations.
It is also stressed that the present FCC implementation is suitable for introducing screenings exploiting the sparsity of the exponential ansatz compared to the previous higher-order CC using intermediates.

\begin{figure}[t]
\includegraphics[width=250pt]{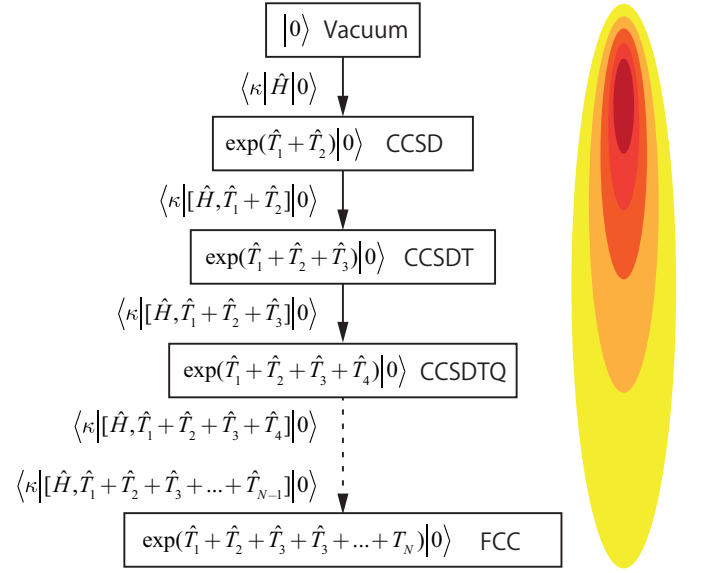}
\caption{Pictorial representation of the iterative expansion of the excitation manifold. The hierarchy of the projection manifold towards FCC is formulated starting from the fermi vacuum.}
\label{fig:exm}
\end{figure}

The FCCR approach is detailed below:
The first-order interacting space of the Fermi vacuum spans the singles and doubles (SD) for the CCSD model.
Then, only the one-rank higher excitation manifold ({\it e.g.}, the triples space with respect to CCSD) is incorporated by taking the single commutator $[\hat H,\hat T]$ using $\hat T$ converged in the proceeding CC calculation.
This update is successively continued until the entire Hilbert space for FCC is integrated.
In this case, the great majority of cluster operators $\hat T_\lambda$ in FCC would possess nearly-null amplitudes unlike those in FCI.
Accordingly, the corresponding excitations connected to these operators are unwanted from the projection manifold.
A necessary modification to the update in FIG. \ref{fig:exm} from this sparsity is to expand the excitation manifold using the {\it primary} set of cluster operators with absolute amplitudes exceeding the connectivity screening threshold, $\vartheta_{\rm C}$, as
\begin{align}
\hat T_{\kappa} \Leftarrow \langle \kappa |[\hat H,\hat T_{\lambda}]|0 \rangle \quad \forall |t_{\lambda}| > \vartheta_{\rm C}.
\end{align}
The excitation manifold of FCCR is formed as a subspace of the FCC model discarding the space connected with nearly-null clusters, and FCCR reduces to FCC in the limit, $\vartheta_{\rm C}=0$.
The connectivity through the higher-order commutators is suitably incorporated by iteratively updating the excitation manifold applying the single commutators.
Incidentally, it turned out that the use of the single commutator for the same purpose was independently conceived by Evangelista \footnote{F. A. Evangelista, private communication (2018).}.

For cluster operators in the space generated by the method described above, the attenuation in the amplitude of the disconnected products further allows us to reduce the operations of the commutators in Eq. (\ref{eq:fcc_w}).
Due to the nonlinear nature of the exponential ansatz, the greater part of the commutator contributions are negligible even after the formation of the FCCR excitation manifold.
Accordingly, the working equation is modified as
\begin{align}
\langle \kappa | e^{-\hat T} \hat H e^{\hat T} | 0 \rangle \approx \langle \kappa | \hat H_{\kappa} | 0 \rangle + {\sum\limits_{\lambda \ne \kappa}^{|t_{\lambda}| > \vartheta_{\rm O}} {\langle \kappa | [\hat H_{\kappa},\hat T_{\lambda}] | 0 \rangle}} \nonumber \\
+ \frac{1}{2}{\sum\limits_{(\lambda\mu) \ne \kappa}^{|t_{\lambda}t_{\mu}| > \vartheta_{\rm O}} {\langle \kappa | [[\hat H_{\kappa}, \hat T_{\lambda}],\hat T_{\mu}] | 0 \rangle}} \nonumber \\
+ \frac{1}{6}{\sum\limits_{(\lambda\mu\nu) \ne \kappa}^{|t_{\lambda}t_{\mu}t_{\nu}| > \vartheta_{\rm O}} {\langle \kappa | [[[\hat H_{\kappa}, \hat T_{\lambda}],\hat T_{\mu}],\hat T_{\nu}] | 0 \rangle}} \nonumber \\
+ \frac{1}{24}{\sum\limits_{(\lambda \mu \nu o) \ne \kappa}^{|t_{\lambda}t_{\mu}t_{\nu}t_{o}| > \vartheta_{\rm O}} {\langle \kappa | [[[[\hat H, \hat T_{\lambda}],\hat T_{\mu}],\hat T_{\nu}],\hat T_{o}] | 0 \rangle}}=0, \label{eq:fccr_w}
\end{align}
such that the commutators with small amplitude are discarded using the operation screening threshold $\vartheta_{\rm O}$, where it is defined that $\hat H_{\kappa} = \exp(-\hat T_{\kappa}) \hat H \exp(\hat T_{\kappa})$, whose triple and quadruple commutator contributions to (\ref{eq:fccr_w}) are null due to the coincident excitation indices, and the limitation in the summation, $(\dots)\ne\kappa$, means none of the indices in parentheses takes the value of $\kappa$.
Eq. (\ref{eq:fccr_w}) becomes exact in the limit $\vartheta_{\rm O} \to 0$.
Importantly, most of the terms with finite power in $T_{\kappa}$ are EPV, and the summation is preserved irrespective of the amplitude $t_{\kappa}$.
This EPV form of the working equation is significant in improving the convergence with respect to $\vartheta_{\rm O}$ compared to the expansion in the non-EPV form, {\it i.e.} the summation is performed for {\it e.g.} $|t_{\kappa} t_{\lambda}| > \vartheta_{\rm O}$.

\begin{figure}[tb]
\includegraphics[width=250pt]{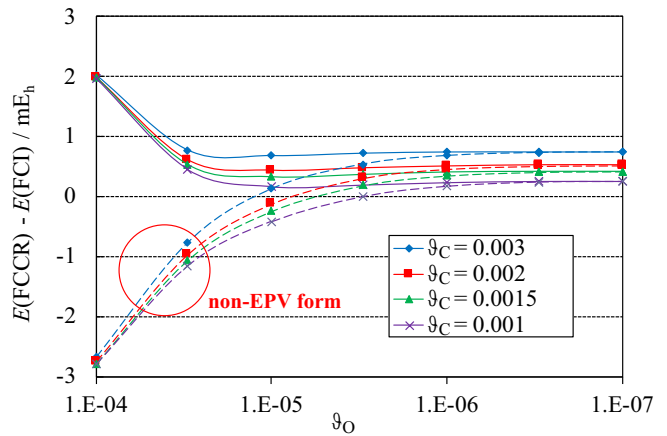}
\caption{Convergence of FCCR correlation energies for N$_2$ with respect to the connectivity and operation screening thresholds, $\vartheta_{\rm C}$ and $\vartheta_{\rm O}$, in cc-pVDZ basis set at the bond distance 3.0 $a_0$. The FCI dimension is $5.4\times10^8$ with 1$s$ electron frozen. The solid and dashed lines denote the results in the EPV and non-EPV forms, respectively.}
\label{fig:epv}
\end{figure}

The efficiency of the FCCR approach is now examined using small molecules in which both dynamic and nondynamic correlation effects are important.
The excitation manifold is updated iteratively using $\vartheta_{\rm O}=\min(4\times10^{-4},\vartheta_{\rm C})$ for a given connectivity screening threshold $\vartheta_{\rm C}$ until the subsequent CC energy difference becomes less than 0.1m$E_h$ for the duration of this study, and then $\vartheta_{\rm O}$ is reduced for the refinement of the energy. 
FIG. \ref{fig:epv} shows the convergence of the FCCR correlation energies for stretched N$_2$ by changing $\vartheta_{\rm C}$ and $\vartheta_{\rm O}$ around the most difficult bond distance.
The number of induced cluster amplitudes increases from 2.2$\times10^5$ to 6.4$\times10^5$ with tightening $\vartheta_{\rm C}$ in the update in the EPV form.
With respect to $\vartheta_{\rm O}$, the FCCR energies tend to converge from below when the EPV form is not applied.
This can be attributed to surplus screening in the quadratic terms compared to the single commutators.
Amongst the quadratic terms, the contribution of EPV is usually quite large as known in CEPA \cite{cepa}, and the EPV form indeed  ameliorates the situation satisfactorily exhibiting almost variational convergence.

\begin{figure}[b]
\includegraphics[width=250pt]{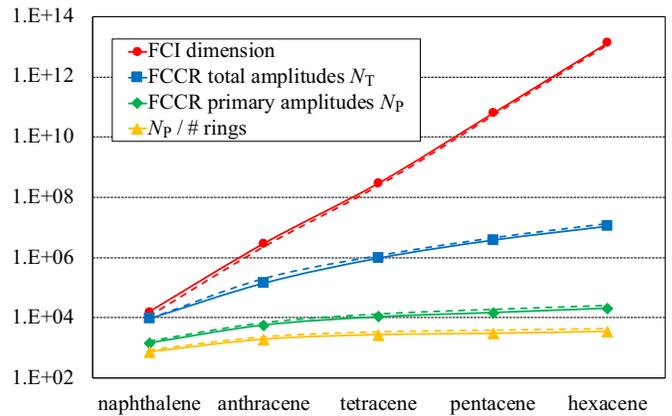}
\caption{The FCI dimension, generated numbers of total and primary amplitudes, $N_{T}$ and $N_{P}$, and $N_{P}$ divided by the number of fused rings of acenes with $\vartheta_{\rm C}=5.0\times10^{-4}$. The solid and dashed lines denote the results for singlet and triplet states, respectively. The $\pi$ orbitals in the STO-3G basis are used to construct the FCI space based on the restricted Hartree-Fock (RHF) canonical orbitals optimized for the singlet states. The Fermi vacuum for a triplet state is generated by the minimum energy single electron excitation from the RHF determinant.
The geometrical parameters are taken from Ref. \onlinecite{hachmann}.}
\label{fig:acene}
\end{figure}

\begin{table*}
\caption[]{\parbox[t]{140mm}{
The deviations of the FCCR singlet and triplet energies ($mE_{\rm h}$) with respect to DMRG \cite{hachmann}, $\Delta E_{\rm FCCR}=E_{\rm FCCR}-E_{\rm DMRG}$, and the arising splittings $\Delta_{\rm ST}$ (kcal/mol) for the model acenes. The operation thresholds are defined as $\vartheta_{\rm O}^{\rm Loose}=3\times10^{-4}$, $\vartheta_{\rm O}^{\rm Middle}=3\times10^{-5}$, and $\vartheta_{\rm O}^{\rm Tight}=3\times10^{-6}$.}}
\begin{tabular}{lrccccccccccccc}
\hline
&& \multicolumn{3}{c}{$\Delta E_{\rm FCCR}$(Singlet)} && \multicolumn{3}{c}{$\Delta E_{\rm FCCR}$(Triplet)} && \multicolumn{3}{c}{$\Delta_{\rm ST}$(FCCR)} & \multicolumn{1}{c}{$\Delta_{\rm ST}$(DMRG)}\\
\cline{3-5} \cline{7-9} \cline{11-13}
&& $\vartheta_{\rm O}^{\rm Loose}$ & $\vartheta_{\rm O}^{\rm Middle}$ & $\vartheta_{\rm O}^{\rm Tight}$ && $\vartheta_{\rm O}^{\rm Loose}$ & $\vartheta_{\rm O}^{\rm Middle}$ & $\vartheta_{\rm O}^{\rm Tight}$ && $\vartheta_{\rm O}^{\rm Loose}$ & $\vartheta_{\rm O}^{\rm Middle}$ & $\vartheta_{\rm O}^{\rm Tight}$ & \rule[0mm]{0mm}{4mm} \\
\hline
naphtalene && 0.3 & 0.0 & 0.1 && 0.2 & 0.0 & 0.0 && 61.5 & 61.5 & 61.5 & 61.5 \\
anthracene && 2.1 & 0.1 & 0.1 && 2.1 & 0.1 & 0.0 && 45.9 & 46.0 & 45.9 & 45.9 \\
tetracene && 7.6 & 0.4 & 0.1 && 7.2 & 0.7 & 0.2 && 34.6 & 35.0 & 34.8 & 34.7 \\
pentacene && 16.4 & 0.9 & 0.1 && 16.0 & 1.8 & 0.4 && 26.4 & 27.3 & 26.9 & 26.7 \\
hexacene && 26.7 & 1.8 & 0.0 && 26.9 & 3.2 & 0.5 && 21.1 & 21.9 & 21.3 & 21.0 \\
\hline
\end{tabular}\label{tab:acene}
\end{table*}

The next example is the singlet-triplet splitting of acenes.
FIG. \ref{fig:acene} shows the growth in the numbers of the generated total and primary amplitudes, $N_{T}$ and $N_{P}$ with increasing the number of fused benzene rings.
The increase in $N_{T}$ is much milder than in the FCI dimension, {\it e.g.} the numbers of the generated FCCR amplitude is 6 order of magnitude smaller than the FCI dimension for hexacene both for the singlet and triplet states.
The lines for $N_{\rm P}$ over the number of rings (in yellow) are almost flat beyond anthracene, indicating that the primary amplitudes increases only linearly with the system size both for singlet and triplet.
Table \ref{tab:acene} lists the errors in the FCCR energies and singlet-triplet splittings with different $\vartheta_{\rm O}$ using the same excitation manifolds (see supplementary material for the full list of total energies \footnote{supplementary material}).
The errors due to the operation screening, which increase with the number of rings especially with the loose screening threshold, $\vartheta_{\rm O}^{\rm Loose}$, tends to cancel between the singlet and triplet states leading to an accurate $\Delta_{\rm ST}$ in the entire range of $\vartheta_{\rm O}$.
The present FCCR manifolds of $\vartheta_{\rm C}=5.0\times10^{-4}$ provide quite accurate results both for the energies and gaps, and the largest error in the total energy with $\vartheta_{\rm O}^{\rm Tight}$ is 0.5 $mE_{\rm h}$ for the triplet state of hexacene.

\begin{center}
\begin{table}
\caption{The number of primary and total amplitudes, $N_{\rm P}$ and $N_{\rm T}$, respectively, and the state energy of FCCR for Cr$_2$ in $E_{\rm h}$ compared to truncated CC and DMRG \cite{dmrg_cr2} correlating 24 electrons in 30 RHF orbitals of the SV basis set of Ahlrichs \cite{svp} with different connectivity thresholds. The tighter operation screening threshold, $\vartheta_{\rm O}=3\times10^{-7}$, is employed for this system, which shows slow and non-variational convergence with respect to $\vartheta_{\rm O}$.}
\begin{tabular}{llrrrrrl}
\hline
& \multicolumn{1}{c}{$\vartheta_{\rm C}$} && \multicolumn{1}{c}{$N_{\rm P}(M)$\footnote{The number of renormalized states $M$ for DMRG.}} && \multicolumn{1}{c}{$N_{\rm T}$
} && \multicolumn{1}{c}{$E$} \\
\hline
FCCR &$6.0\times10^{-4}$ && 11,399 && 8,909,199 && -2086.4159 \\
&$4.0\times10^{-4}$ && 19,244 && 15,162,543 && -2086.4169 \\
&$2.0\times10^{-4}$ && 42,813 && 32,043,659 &&  -2086.4186 \\
&$1.0\times10^{-4}$ && 95,849 && 68,766,328 && -2086.4203 \\
\hline
CCSD &&&&& 8,766 && -2086.3225 \\
CCSDT &&&&& 598,082 && -2086.3805 \\
CCSDTQ &&&&& 23,422,496 && -2086.4067 \\
CCSDTQP &&&&& 560,106,440 && -2086.4144 \\
\hline
DMRG &&&2,000 &&&& -2086.4198 \\
&&& 5,000 &&&& -2086.4206 \\
&&& 10,000 &&&& -2086.4208 \\
Extrapolated &&& $\infty$ &&&& -2086.4210 \\
\hline
\end{tabular}\label{tab:cr2}
\end{table}
\end{center}

Finally, the convergence of the total energy of Cr$_2$ at a bond length of 1.5 \text{\AA} is shown in Table \ref{tab:cr2}.
This system is very slow in the convergence of excitation ranks, and non-negligible septuple and octuple excitations appear in accordance with the observations of Lehtola {et al.} \cite{cdfci}.
The energies of the truncated CC are higher than those of FCCR even for the CCSDTQP model with 560 million amplitudes due to the absence of these high-rank cluster operators.    
With tightening $\vartheta_{\rm C}$, the FCCR energy decreases, and the case employing 69 million amplitudes gives -2086.4203 $E_{\rm h}$, that is only 0.7 $mE_{\rm h}$ higher than the best extrapolated DMRG estimate of -2086.4210 $E_{\rm h}$ \cite{dmrg_cr2}.
The complementary space connected to the secondary cluster operators is not used in the present FCCR calculations, and it is likely that a perturbative correction to the space mitigates this small discrepancy.

In conclusion, FCCR has been introduced based on an FCC of direct commutator expansions.
The sparsity of the cluster operators facilitates efficient screenings for the excitation manifold and commutator operations.
Although high-rank cluster operators are needed for strong electron correlation, the treatment is feasible provided the amplitudes are not prohibitively numerous.
Therefore, FCCR appears to be a promising means for strongly correlated electronic systems in a balanced descriptions of dynamic and non-dynamic correlation effects.
Note that FCCR can also be implemented in terms of the Wick theorem instead of the direct commutator expansions, {\it i.e.}, first generate all possible connected diagrams from $\hat H$ and $\hat T_{\lambda},\hat T_{\mu},\dots$ and then seek for the interacting states $\langle \kappa |$.
In that case, an automatic code synthesis is likely to be utilized for a diagrammatic implementation of FCCR.
With regard to EPV, Bartlett and Musia{\l} \cite{ncc} have developed an $n$CC hierarchy, and a selected model in conjunction to $n$CC appears to be an interesting direction as well.
In addition, various extensions of FCCR can be envisaged, {\it e.g.}, (i) acceleration of convergence using a different choice of Fermi vacuum including Bruecknerization, (ii) orbital rotations in the occupied and virtual spaces, (iii) perturbative corrections with respect to the complementary space connected to the secondary clusters, and so on.
This line of research, with applications including more challenging systems, will be pursued and reported in the near future.

The authors thank Takashi Tsuchimochi for helpful comments on the manuscript.
This work was partially supported by MEXT as Priority Issue on Post-K computer" (Development of new fundamental technologies for high-efficiency energy creation, conversion/storage and use) using the computational resources of the K computer provided by the RIKEN Advanced Institute for Computational Science through the HPCI System Research project (Project ID: hp160202, hp170259, hp180216) and JSPS Grant-in-Aids for Scientific Research (A) (Grant No. JP18H03900).

\nocite{*}

\bibliography{refs}

\end{document}